\documentclass[aps,prb,twocolumn,groupedaddress,showpacs]{revtex4-1}

\usepackage{amsmath}
\usepackage{graphicx}
\usepackage{dcolumn}

\begin{document}
\title{Coherent quantum transport of charge density waves}
\author{J.~H.~Miller,~Jr.}
\email{jhmiller@uh.edu}
\affiliation{Department of Physics, University of Houston, Houston, Texas 77204-5005 USA}
\affiliation{Texas Center for Superconductivity, University of Houston, Houston, Texas 77204-5002 USA}
\author{A.~I.~Wijesinghe}
\email{aiwijesinghe@yahoo.com}
\affiliation{Department of Physics, University of Houston, Houston, Texas 77204-5005 USA} 
\affiliation{Texas Center for Superconductivity, University of Houston, Houston, Texas 77204-5002 USA}

\author{Z.~Tang}
\affiliation{Department of Chemistry, University of Houston, Houston, Texas 77204-5003 USA}
\author{A.~M.~Guloy}
\affiliation{Texas Center for Superconductivity, University of Houston, Houston, Texas 77204-5002 USA}
\affiliation{Department of Chemistry, University of Houston, Houston, Texas 77204-5003 USA}

\date{\today}

\begin{abstract}
Recent experiments show oscillations of dominant period $h/2e$ in conductance \textit{vs}. magnetic flux of charge density wave (CDW) rings above 77~K, revealing macroscopically observable quantum behavior. The time-correlated soliton tunneling model discussed here is based on coherent, Josephson-like tunneling of microscopic quantum solitons of charge $2e$. The model interprets the CDW threshold electric field as a Coulomb blockade threshold for soliton pair creation, often much smaller than the classical depinning field but with the same impurity dependence (e.g., $\sim n_i^2$ for weak pinning). This picture draws upon the theory of time-correlated single-electron tunneling to interpret CDW dynamics above threshold. Similar to Feynman's derivation of the Josephson current-phase relation for a superconducting tunnel junction, the picture treats the \textit{Schr\"{o}dinger equation as an emergent classical equation} to describe the time-evolution of Josephson-coupled  order parameters related to soliton dislocation droplets.  Vector or time-varying scalar potentials can affect the order parameter phases to enable magnetic quantum interference in CDW rings or lead to interesting behavior in response to oscillatory electric fields. The ability to vary both magnitudes and phases is an aspect important to future applications in quantum computing. Published in Physical Review B 87 (11), p. 115127 (2013).
\end{abstract}


\maketitle
\section{Introduction}
Recent developments necessitate a transformation in our understanding of charge density wave (CDW) transport to one based on quantum principles.~\cite{1} The CDW is a correlated electron (or electron-phonon) system that, like a superconductor, can transport electrons through a quasi-one-dimensional or layered crystal \textit{en masse}.~\cite{2,3} It is the only known such system capable, in the linear chain compound NbS$_3$, of collectively carrying electric current above 37$^\circ$C, the temperature of the human body.~\cite{4}  Moreover, a significant body of evidence highlights the importance of CDW,~\cite{5,6} stripe,~\cite{7,8,9} and other charge- and/or spin-ordered phases in high-$T_c$ and other unconventional superconductors as carrier concentration is varied, e.g., by doping. Some experiments~\cite{10} suggest possible interfacial superconductivity or a related phase transition near the boundary between ion-implanted and unimplanted regions of a CDW in NbSe$_3$.

CDW electron wavefunctions are delocalized over long distances, and the charge modulation results from quantum interference between right- and left-moving electron states separated by the nesting wavevector $2k_F$.  The CDW electron condensate, coupled to the $2k_F$ phonon condensate, can thus be viewed as a sticky quantum fluid (or deformable quantum solid with dislocations~\cite{11}) within which microscopic entities can tunnel coherently in a Josephson-like manner, flowing through a barrier like water dripping from a faucet. The jerky current flow in this collective version of time-correlated single electron tunneling~\cite{12, 13} results from the Coulomb blockade effect created by charged CDW phase kinks. If interchain interactions in a linear chain compound are not too strong, then condensed electrons or quantum solitons~\cite{14} in a CDW may be no more impeded from quantum tunneling through a miniscule barrier or pinning gap~\cite{15, 16} than the photons from a laser pointer would be impeded by their large numbers from evanescently decaying through a thin metal film. Here we stress that coherent Josephson-like tunneling of microscopic entities within a condensate is quite different from macroscopic quantum tunneling, despite the misleading titles of some early papers.~\cite{17,18}

The ability to interpret some CDW transport phenomena classically~\cite{2,3} does not imply a need to reject underlying quantum mechanisms, given the fact that electrons behave quantum mechanically. For example, a classical sliding electron theory could have been proposed in the 1890's for electrons flowing through a wire, since Ohm's law is consistent with a linear velocity-force relation.
 Nevertheless if physicists had clung to such a hypothesis, declaring electron transport a ``solved problem,'' any further progress in understanding the behavior of electrons in solids would have halted in its tracks. Coherent Josephson tunneling of electron pairs is another example,~\cite{19} in which  the quantum-mechanical phase difference across the junction is  treated  as a classical variable. Feynman (vol.~III, Ch.~21 of ref.~\cite{20}) provides an elegant derivation of the Josephson current-phase relation by treating the time-dependent Schr\"{o}dinger equation itself as a ``classical'' equation for the coupled superconducting order parameters.

Recent evidence supporting quantum behavior of CDWs includes Aharonov-Bohm (A-B) quantum interference effects in TaS$_3$ rings up to 79~K, showing oscillations with a dominant period of $h/2e$ in CDW conductance \textit{vs.} magnetic flux.~\cite{21} Similar oscillations have been reproduced, as reported in 2012,~\cite{22} for at least five TaS$_3$ rings with circumferences of up to 85~$\mu$m. This size is nearly two orders of magnitude larger than that of typical normal metal rings exhibiting the A-B effect, usually below 1~K.~\cite{23}  The magneto-conductance oscillations are only observed in the CDW, not normal electron, magneto-conductance above the threshold electric field for CDW transport, and the amplitude of the oscillations scales with CDW current. The ring experiments show that, at least for these materials, the CDW condensate exhibits quantum phase coherence over several ten's of microns. Such extraordinary behavior, which manifests Planck's constant at the macroscopic scale, underscores the need for a fundamental paradigm shift in which the laws of quantum physics play a crucial role in describing CDW electron transport.

One ring was reported~\cite{22} to exhibit  telegraph-like temporal switching between high and low CDW current states, the high current state showing substantially larger amplitude A-B oscillations than the low current state. This telegraph-like near destruction and reappearance of A-B quantum interference in the CDW ring indicates quasi-periodic partial loss of quantum coherence, suggesting two types of transport involving either probabilities or probability amplitudes, the former lacking \textit{vs.} the latter including quantum coherence. Alternatively, the behavior may suggest a phase slip process~\cite{22} or perhaps even some form of macroscopically observable wave-function collapse. A reversal in phase of width-peak product \textit{vs.} flux (Fig.~\ref{fig:Fig4}(d) of~\cite{22}) for the two current states shows similarity to switching effects reported in A-B interferometers with embedded Coulomb-blockade quantum dots.~\cite{24} Regardless of which interpretation ultimately emerges, a deeper understanding of the observed behavior based on the laws of quantum physics could ultimately prove important to condensed matter physics and possibly to the foundations of quantum physics.

Given that the CDW order parameter depicts an electron-hole condensate rather than an electron pair condensate as in a superconductor, an important question is whether and, if so, why the predominant period ought to be $h/2e$, also reported in previous A-B experiments on NbSe$_3$ with columnar defects.~\cite{25} An interpretation  in section~\ref{sec:4} suggests that nucleated quantum solitons, of charge $\pm 2e$ per chain for a fully condensed system, quantum-mechanically interfere with themselves around the two branches of the ring. It is stressed, however, that a more realistic model should incorporate disorder to be consistent with the observed  $\sim 10\%$ modulation amplitude and somewhat disordered behavior in the magneto-conductance oscillations. Moreover, the original Aharonov-Bohm paper~\cite{26} proposed quantum interference due to both a magnetic vector potential and a time-varying scalar potential as the charged particle traverses the two branches of a ring. The latter effect, sometimes called the scalar A-B effect, can combine with the magnetic A-B effect to exhibit quantum interference that depends on both voltage and magnetic flux.~\cite{27, 28}  CDWs are important in this regard since, unlike superconducting or normal metal rings with ballistic transport, A-B interference occurs with significant voltage drop (up to 300~mV reported ~\cite{22}) between contacts, showing significant variation in peak amplitudes \textit{vs.} voltage. Some experiments, discussed in section~\ref{sec:4}, involve the application of time-varying voltages that can couple to quantum-mechanical phase in a fashion similar to the scalar A-B effect.

Any viable quantum picture must also explain the threshold electric field for CDW transport, as well as narrow-band noise, coherent voltage oscillations, etc.~\cite{2,3}  NbSe$_3$ and related materials have threshold fields that scale with impurity concentration $n_i$ either as $n_i^2$ (weak pinning) or $n_i$ (strong pinning),~\cite{2,3} depending on sample, consistent with the classical Fukuyama-Lee-Rice (FLR) model~\cite{29,30} of CDW pinning. Early proposals for tunneling of CDW electrons~\cite{15} or solitons~\cite{31,32} lacked compelling interpretations for the threshold field and other phenomena, although Bardeen proposed phenomenological,~\cite{16} sometimes semi-classical~\cite{33} interpretations for the threshold field and narrow-band noise. However, a key paper on the quantum picture emerged in 1985,~\cite{34} pointing out that nucleated solitons and antisolitons of charge $\pm q$ generate an internal field $E^*\propto q/\epsilon$, whose electrostatic energy $\frac{1}{2} \epsilon E^{*2}$ prevents soliton tunneling for applied fields less than a threshold $E_T=E^*/2$ without violating energy conservation. Critically, although this Coulomb blockade threshold can be much smaller than the classical depinning field, it exhibits the same scaling with impurity concentration. This is because the CDW's polarizability and dielectric response $\epsilon$ vary inversely with pinning strength, as further discussed in section~\ref{sec:2}, which also discusses the possible existence of both Coulomb blockade and classical depinning fields in some materials.

\begin{figure}[h]\centering		
\includegraphics[scale=.0825]{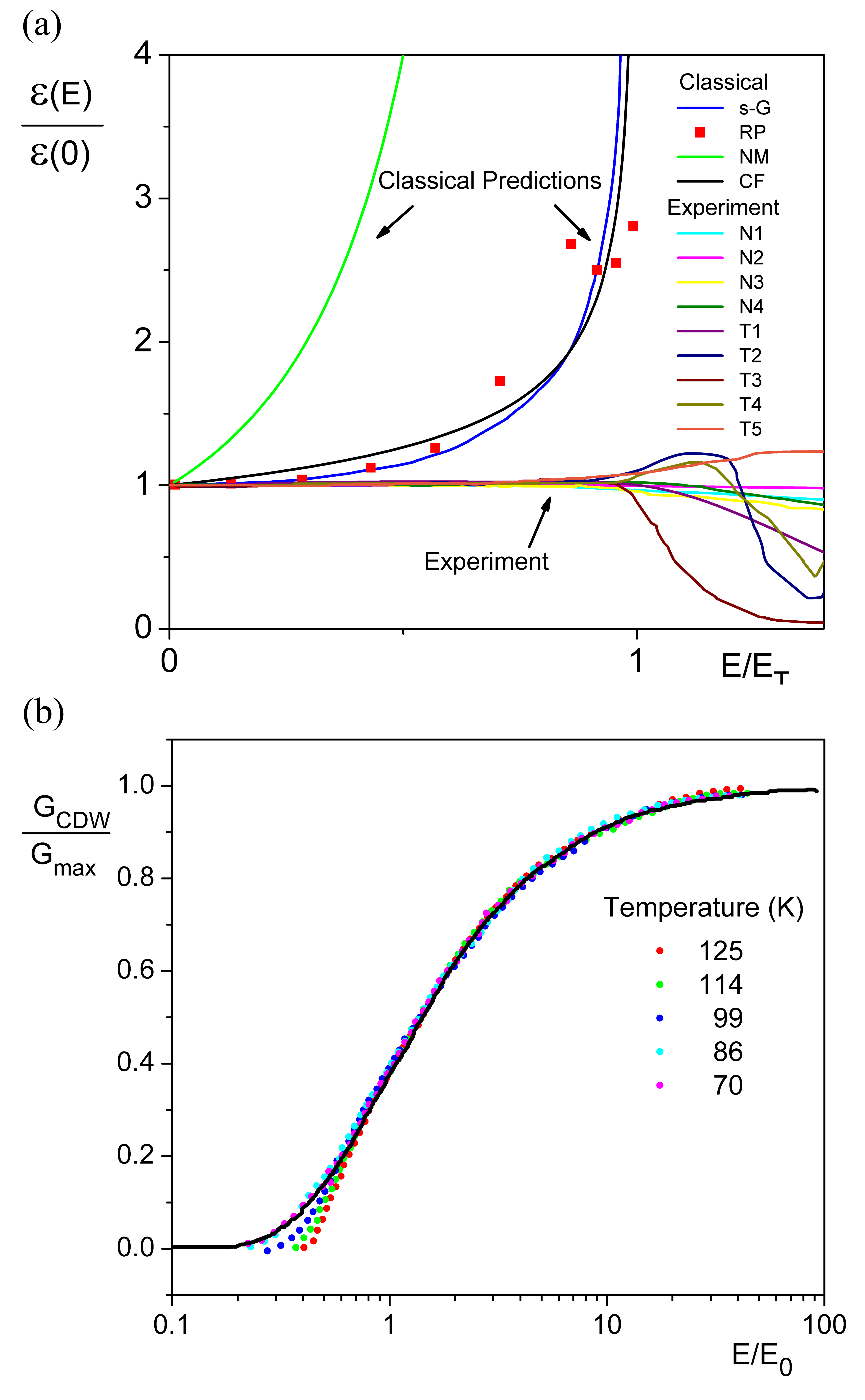}
\caption{(a)~Bias dependent dielectric response, showing classical predictions \textit{vs.} experiment. Classical models include classical sine-Gordon (s-G); random pinning (RP);~\cite{41} renormalization group (NM)~\cite{42} $\left|f\right|^{-2}$; and incommensurate harmonic chain (CF),~\cite{43} $\left|f\right|^{-0.34}$,  models,  where $f = 1-E/E_T$. Some NbSe$_3$ measurements were carried out in our lab using a bridge circuit (NbSe$_3$: \underline{N1}, 45~K, 1~kHz; \underline{N2}, 120~K, 3~kHz; \underline{N3}, 45~K, 100~kHz),  while additional measurements were carried out by ZG~\cite{44} (\underline{N4}, NbSe$_3$, 42~K, 3.2~MHz)  and WMG~\cite{45} (TaS$_3$: \underline{T1}, 130~K, 5~MHz; \underline{T2}, 100~K, 1~kHz; \underline{T3}, 110~K, 1~kHz; \underline{T4}, 100~K, 1~kHz; \underline{T5}, 100~K, 10~kHz). (b)~Experimental CDW conductance \textit{vs.} electric field for NbSe$_3$ as compared to the Zener tunneling curve $\exp[-E_0/E]$ (solid line) pointed out by Bardeen.~\cite{46} Adapted with permission from,~\cite{46} Copyright 1990, American Institute of Physics.} 
\label{fig:Fig1}
\end{figure}

Several experiments indicate that, in NbSe$_3$ and orthorhombic TaS$_3$, the CDW displaces very little below threshold suggesting that, in these materials, the measured threshold is the Coulomb blockade threshold rather than the classical depinning field. This is evident in NMR experiments~\cite{35} showing a 2$^\circ$ CDW phase displacement in NbSe$_3$, as compared to the classically predicted 90$^\circ$ displacement just below threshold. Further evidence is provided by dielectric and other ac response (mixing, etc.~\cite{36,37,38}) measurements, which exhibit a flat bias dependence as compared to the classically predicted divergent dielectric response shown in Fig.~\ref{fig:Fig1}(a). These experiments reveal that, even just below threshold, each portion of the CDW sits near the bottom of a pinning potential well. This suggests that, at least for these samples, the measured threshold is substantially smaller than the classical depinning field and likely a Coulomb blockade threshold for charge soliton nucleation.~\cite{1,34,39,40}

Bardeen's model of coherent Zener tunneling of CDW electrons through a tiny pinning gap~\cite{15,16} fixed at $\pm k_F$, unlike the Peierls gaps which can displace in momentum space, was motivated by the shape of the \textit{I-V} characteristic. This has been found, starting with the early experiments on NbSe$_3$,~\cite{47} to progress from a rounded Zener tunneling-like characteristic~\cite{46} (Fig.~\ref{fig:Fig1}(b)) to a nearly piecewise linear form in crystals with fewer impurities. This behavior is consistant with soliton pair creation with a Coulomb blockade threshold.~\cite{1,34,40} Soliton pair creation is analogous to Landau-Zener tunneling, recently applied to Fermi superfluid gases,~\cite{48} Schwinger pair production,~\cite{49,50} or creation of superconducting vortex-antivortex pairs.~\cite{51} The existing classical models fail to explain the shape of the CDW \textit{I-V} curves of NbSe$_3$ and orthorhombic TaS$_3$ in a straightforward fashion.~\cite{52} Moreover, there is no compelling classical sliding interpretation for the  quantum interference effects seen in CDW rings.~\cite{21,22} Any viable CDW transport theory of this extraordinary phenomenon must contain Planck's constant, even at the macroscopic level. However, this does not rule out the possibility of using the Schr\"{o}dinger equation itself as an emergent `classical' equation, as discussed by Feynman in the context of superconductivity (ref.~\cite{20}, vol.~III, Ch.~21). This approach, novel for CDWs,~\cite{1} of employing the Schr\"{o}dinger equation to describe classically robust complex order parameters related to soliton dislocation droplets, will be discussed in section~\ref{sec:3}. The following section discusses a modified sine-Gordon model, the simplest possible model of a pinned CDW.

\section{Pinned charge density wave as massive Schwinger model}\label{sec:2}
A CDW has a modulated charge $\rho(x,t)=\rho_0 (x,t)+\rho_1  \cos[2k_F x-\phi(x,t)]$ along the axis of a linear chain compound. Here $\rho_0 (x,t)$ contains background charge and any excess or deficiency of charge $\propto \partial\phi/\partial x$. The entire CDW condensate and Peierls gaps, initially at $\pm k_F$, can be displaced in momentum space, resulting in a current: $I_{cdw}\propto \partial\phi/\partial t$.~\cite{53,54} Although a real CDW is pinned by impurities, in some materials it will still transport a current provided the applied field $E$ exceeds a threshold $E_T$. Displacing the CDW by one wavelength (advancing $\phi$  by $2\pi$) returns the system to its original state (except for charge displaced between contacts) so the pinning energy is periodic in $\phi$:  $u_p [1-\cos \phi ]$. (A quantum version~\cite{55} of the FLR model,~\cite{29,30} including disorder, would be more accurate but observed voltage oscillations suggest the simple sine-Gordon picture captures much of the physics for high quality crystals.) This simplified picture, resulting from impurities, is similar to that which would result from a commensurability index $M = 1$.

Unlike a superconductor, the CDW charge modulation, whose order parameter corresponds to electron-hole pairing and carries no net charge, does not couple directly to a uniform electric field or vector potential. However, \textit{gradients} or kinks in CDW phase carry charges that (1) couple to an externally applied field and (2) generate their own electric fields that lead to electrostatic interactions. These electrostatic interactions between kinks, often neglected in previous theories, are important whether treating the system classically or quantum mechanically. If the CDW phase is initially fixed at zero at the contacts or at  $\pm \infty$, advancing the phase by  $\phi$ in the middle creates charged kinks that produce an internal field: $E_\phi  = (E^*/2\pi)\phi$ , where $E^*=2e/(\epsilon A_{ch})$ is the field created by a $2\pi$  phase soliton-antisoliton pair and $A_{ch}$ is the cross-sectional area per chain.

\begin{figure}[h]\centering		
\includegraphics[scale=.35]{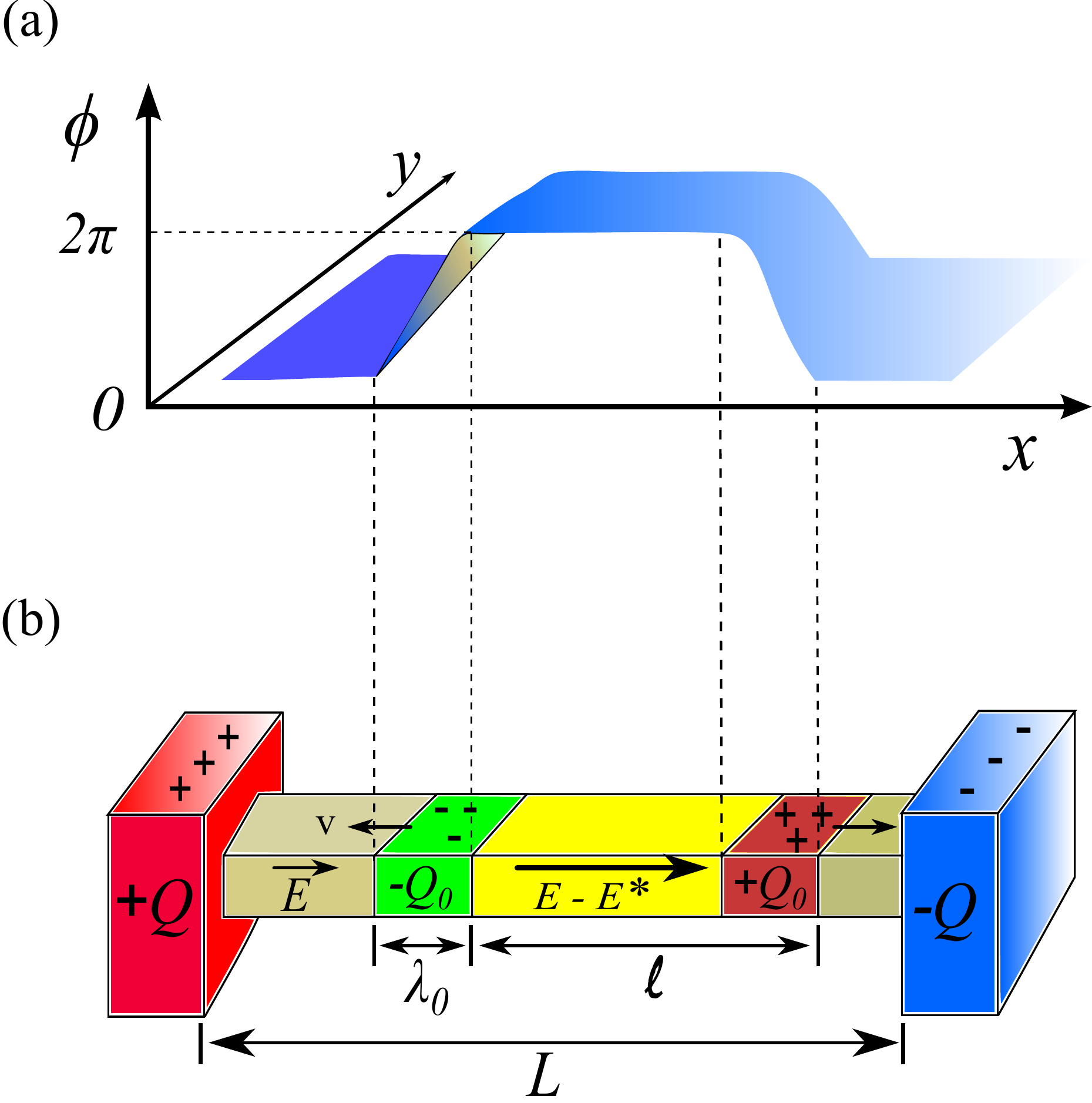}	
\caption{(a)~CDW phase \textit{vs.} position, showing internal field $E^*$ produced by a soliton-antisoliton domain wall pair. (b)~Model of density wave capacitance showing nucleated domain walls, more realistically depicted as soliton dislocation droplets in section~\ref{sec:3}. The applied field $E$ partially or completely cancels the internal field $E^*$.}	
\label{fig:Fig2}
\end{figure}

Figure~\ref{fig:Fig2} shows the combined effects of the applied field $E$ and the field $E^*$ created by a pair of soliton domain walls.  The difference in electrostatic energy densities, $\frac{1}{2}\epsilon (E \pm E^*)^2-\frac{1}{2} \epsilon E^2$, with and without the pair is positive when $E$ is less than the Coulomb blockade threshold field, $E_T=\frac{1}{2}E^*=en_{ch}/\epsilon$. Here $n_{ch}=1/A_{ch}$ is the number of parallel chains per unit cross-sectional area. The empirically observed relation $\epsilon E_T \sim en_{ch}$ pointed out by Gr\"{u}ner~\cite{56,57} thus emerges naturally from this picture. The simplest classical model predicts~\cite{56,57} the classical depinning field $E_{cl}$ to scale as: $\epsilon E_{cl}=4\pi en_{ch}$, where $E_{cl}\propto n_i^2$ for weak pinning ($n_i$ being the impurity concentration). This yields: $E_T=E_{cl}/4\pi$, which has the same impurity dependence as $E_{cl}$ for a fixed temperature. Screening by normal carriers further enhances $\epsilon$ and reduces the ratio: $E_T/E_{cl}$.  For fixed $n_i$, the temperature dependence of carrier concentration and $\epsilon$ leads (inversely) to the strong temperature dependence of $E_T$ seen in some materials. In addition, $en_{ch}$ is multiplied by the condensate fraction  $\rho_c$ in a more precise description.

Normal carrier screening may also allow the modified sine-Gordon (massive Schwinger) model to work in some materials despite the fact that, per FLR,~\cite{29, 30} a real CDW pinned by impurities is expected to be deformed even in its ground state. A static phase kink in the ground state, like a nucleated soliton, carries charge, but the normal carriers have plenty of time to completely screen it out. However, any `bubble' of lower energy nucleated by an applied field, where the phase locally advances by $2\pi$ to a lower pinning potential well, will be bounded by regions that depart from the ground state in such a way that nucleated soliton-like charges will become exposed as the normal electrons take a finite time to respond. Substantial screening even for such transient events, however, will likely still be enough to greatly reduce  $E^*$ and the Coulomb blockade threshold $E_T$.

\begin{figure}[h]	\centering		
\includegraphics[scale=0.4]{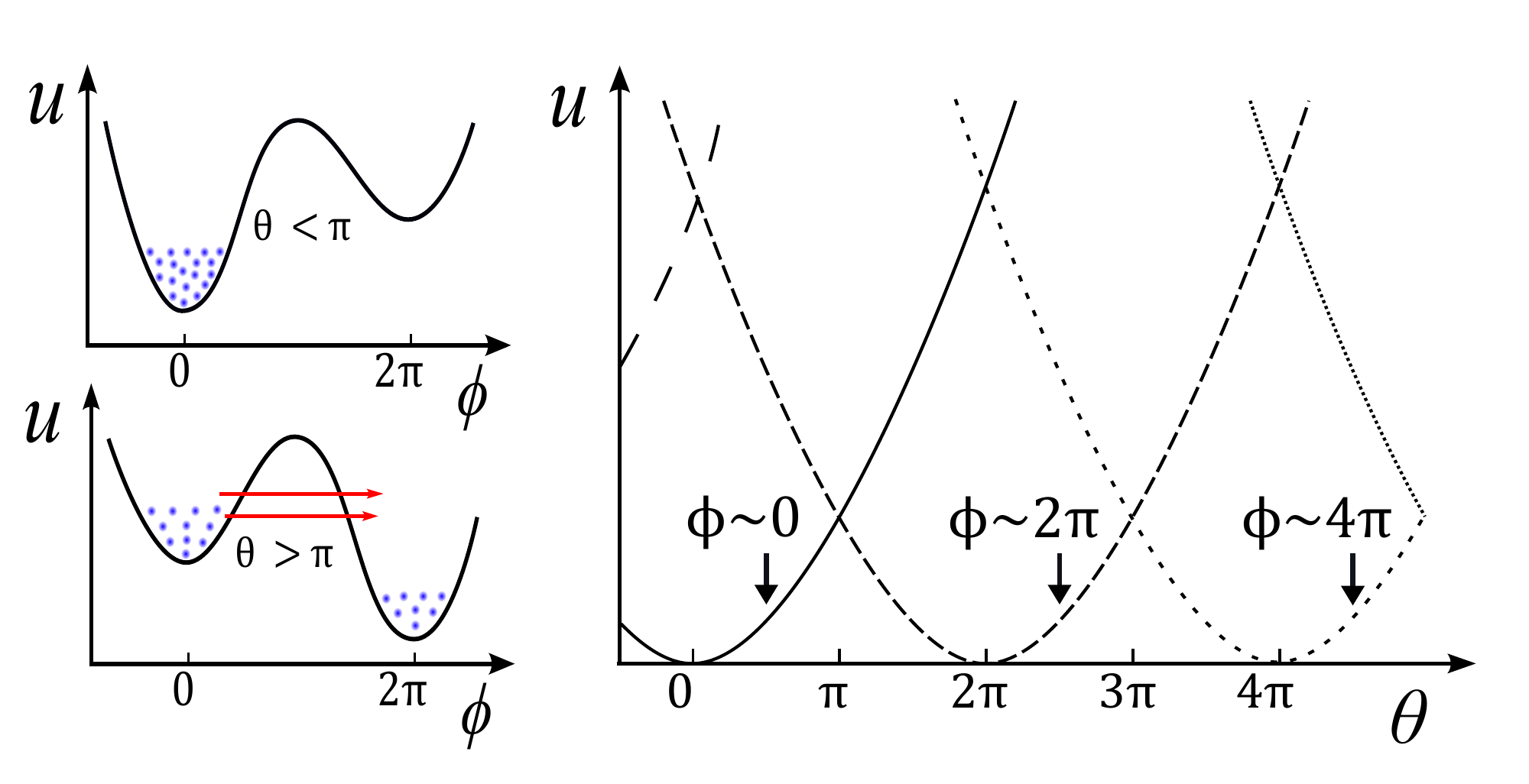}	
\caption{(Left)~Potential energy \textit{vs}. $\phi$  for two values of $\theta$, with many degrees of freedom illustrated as blue dots.  Tunneling can only occur if   $\theta >  \pi  (E > E_T)$, when ``bubbles" of the phases  $\phi_k$ for the parallel CDW chains can nucleate by tunneling into the adjacent well.  (Right)~ Potential energy parabolas $u$ \textit{vs.} $\theta$, in which the phases  $\phi_k \approx \phi$    are sitting in various potential minima,   $\phi \sim 2\pi n$. The first crossover between parabolic branches occurs at $\theta=\pi$.}		
\label{fig:Fig3}
\end{figure}

Following the quantum field theory literature, the applied field $E$ relates to the `vacuum angle' as $\theta = 2\pi(E/E^*)$.  For phase displacements $\phi$  between contacts, $E$ partially cancels $E_\phi$, yielding an electrostatic energy $u_E (\theta-\phi)^2$.~\cite{1}  The potential energy per chain can then be written as:~\cite{1,40,58} 
\begin{equation}
U[\phi]=\int dx\left\{2u_p\left[1-\cos\phi(x)\right]+u_E\left(\theta-\phi(x)\right)^2\right\}.
\label{eq:eq1}
\end{equation}
This is a variant of the bosonic massive Schwinger model, studied as a model of quark confinement~\cite{39,59} and first adapted to explain the CDW quantum threshold field by Krive and Rozhavsky.~\cite{34} The usual linear coupling $\propto -\theta\phi$ is contained in the quadratic term, as are electrostatic contributions $\propto \phi^2$ and $\theta^2$. When $\theta<\pi$, the system is stable classically and quantum mechanically (Fig.~\ref{fig:Fig3}). When $\theta>\pi$, the $\phi \sim 2\pi$ state becomes the lowest energy state.

Thus, $\theta=\pi$ demarcates the boundary~\cite{58} above which the system can decay into the lower well. Several quasi-1-D systems appear to be in the sweet spot of interchain interactions - strong enough to avoid being swamped by thermally excited soliton dislocations but not strong enough to remain forever trapped in the higher well. Some NbSe$_3$ crystals suddenly switch into a higher CDW current-carrying state as the field is increased~\cite{60} and show a hysteretic \textit{I-V} curve. A natural interpretation is that, as  $\theta$ is increased above $\pi$, the system is temporarily trapped in the higher metastable well (Fig.~\ref{fig:Fig3}) before decaying rapidly into the lower well. Other materials show more than one threshold field.~\cite{61,62}  The picture here provides a simple interpretation: that the lower threshold field is the Coulomb blockade threshold for soliton nucleation~\cite{1,34,40}  while the upper threshold is the classical depinning field.

\begin{figure}[h]	\centering		
\includegraphics[scale=0.45]{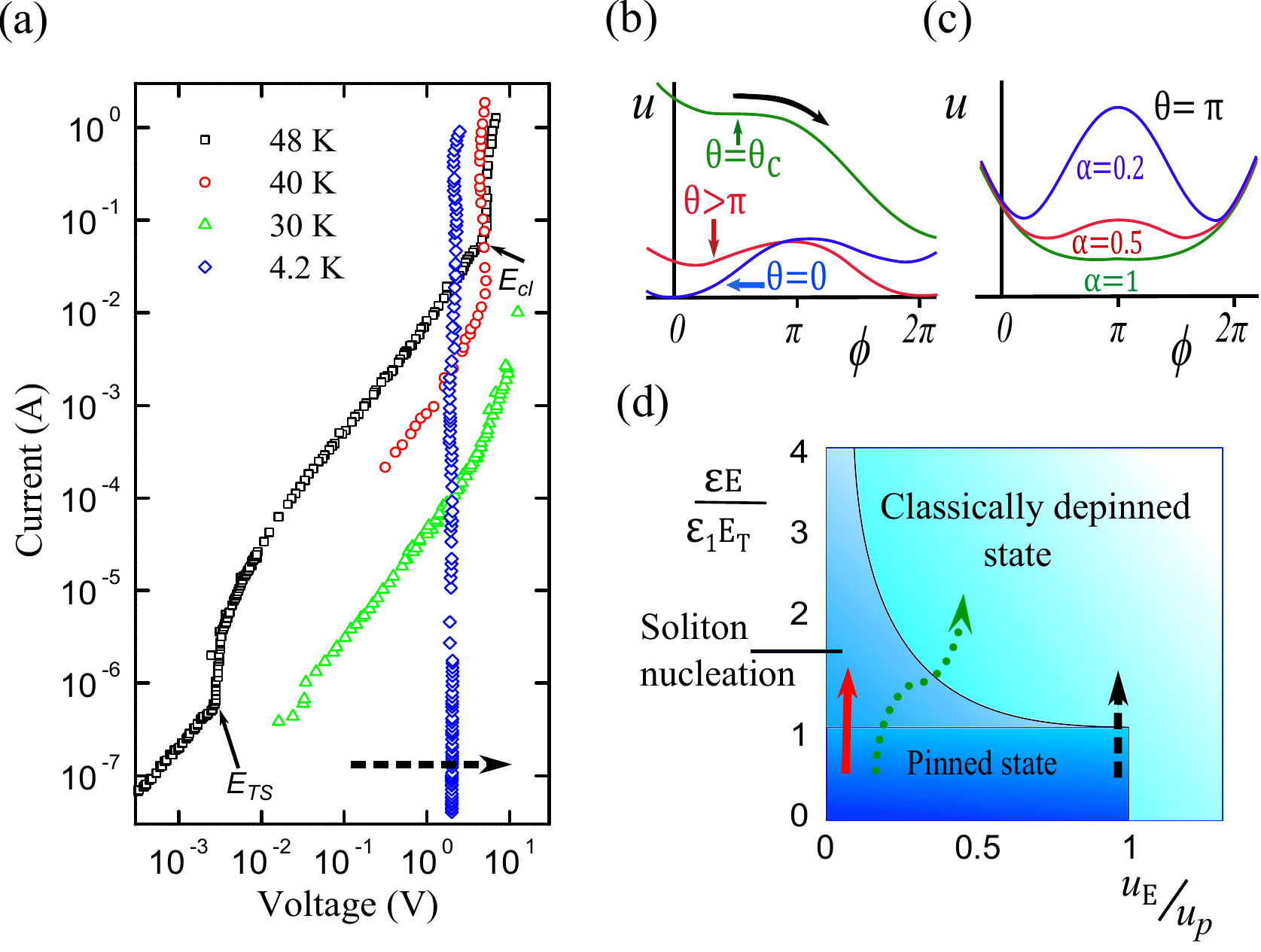}    	
\caption{(a)~Blue bronze \textit{I-V} curves~\cite{62} in which two threshold fields are apparent.  (b)~Potential energy \textit{vs.}  $\phi$, for increasing values of $\theta$  up to the classical depinning instability  $\theta_c$. (c)~$u$ \textit{vs.} $\phi$ for several ratios $u_E/u_p$ when  $ \theta=\pi $.  (d)~Phase diagram~\cite{1} showing pinned, soliton nucleation, and classically depinned states. \underline{Red arrow} ($u_E/ u_p << 1$) crosses from the pinned state into the soliton nucleation region. The \underline{dotted green arrow} depicts a system exhibiting both thresholds. Since $u_E \propto 1/\epsilon$  the path curves to the left (right) if  $\epsilon$ increases (decreases) with field. \underline{Dashed black arrows}: classical depinning dominates, as suggested by the 4.2-K blue bronze data.}	
\label{fig:Fig4}
\end{figure}

Figure~\ref{fig:Fig4}(a) shows blue bronze data~\cite{62} that, especially at 48~K, exhibits two distinct threshold fields above which the conductance increases. The upper threshold field, presumed to be the classical depinning field $E_{cl}$, shows the most dramatic increase in CDW current. The lower threshold field is interpreted as the Coulomb blockade field $E_{Ts}$ for soliton nucleation. Figure~\ref{fig:Fig4}(b) shows plots of $u$ \textit{vs.} $\phi$, illustrating the soliton nucleation ($\theta \geq \pi$) and classical depinning ($\theta \geq \theta_c$) instabilities that arise as  $\theta$ is increased. Figure~\ref{fig:Fig4}(c) plots $u$ \textit{vs.} $\phi$  when $\theta=\pi$ for several values of $u_E/u_p$. Figure~\ref{fig:Fig4}(d) shows the resulting phase diagram,~\cite{1} which plots $\theta/\pi=\epsilon E/\epsilon_1 E_T$ \textit{vs.}  $u_E/u_p$ and allows for variations in $\epsilon$ relative to its threshold value $\epsilon_1$.  The diagram illustrates the pinned state $(\theta<\pi, u_E/u_p <1)$, a region in which soliton nucleation occurs $(\pi<\theta<\theta_c)$, and a classical depinning region $(\theta>\theta_c)$.

The flat dielectric and other ac responses~\cite{36,38} (Fig.~\ref{fig:Fig1}) and small phase displacements~\cite{35} below threshold in NbSe$_3$ and TaS$_3$ suggest $u_E/u_p <<1$ [solid red arrow in Fig.~\ref{fig:Fig4} phase diagram] in these samples. The computed phase displacement $\left\langle \phi\right\rangle$ below threshold~\cite{40} compares favorably to the reported 2$^\circ$ value~\cite{35} for NbSe$_3$ provided $u_E/u_p  \sim 0.015$.~\cite{40} Using  $u_E/u_p =2\pi E_T/E_{cl}$ , the 48-K blue bronze data~\cite{62} in Fig.~\ref{fig:Fig4} suggests a similar value of about 0.01. The increase in $E_{Ts} (\propto 1/\epsilon)$ with decreasing temperature is readily interpreted as due to a reduction in $\epsilon$ as the normal carrier concentration decreases. At 4~K, the normal carriers are largely frozen out, resulting in a relatively low $\epsilon$  and sufficiently high $u_E/u_p$  for classical depinning to dominate [dashed black arrows in Fig.~\ref{fig:Fig4}]. The following sections discuss CDW dynamics above threshold and the issue of quantum coherence, as revealed by CDW ring~\cite{21,22} and other experiments.

\section{Time-correlated soliton tunneling model}\label{sec:3}
A basic premise of this paper is that much of the dynamical behavior of CDWs seen in the highest quality crystals of NbSe$_3$ and related materials can be understood by extending the simple picture discussed above. These phenomena include narrow-band noise with a fundamental frequency that scales with CDW current and a rich spectrum of harmonics, and complete mode-locking with an external \textit{ac} source at high drift frequencies (even when much higher than the dielectric relaxation frequency, in contradiction with classical predictions~\cite{63}). The key to successfully applying such a simple model is to accept quantum principles, one of which is Gell-Mann's totalitarian principle:~\cite{64} ``Everything not forbidden is compulsory." Applied to CDWs the implication is: \textit{If CDW electrons can tunnel then they must tunnel}. Experiments to date suggest that CDW condensates behave as sticky quantum fluids or deformable quantum solids with dislocations~\cite{11} rather than massive classical deformable objects. 

Hypotheses addressed in this paper include: {1)} low energy phase soliton dislocations of charge $\pm 2e$ (or, in our view less likely, amplitude solitons of charge $\pm e$)~\cite{65,66,67,68,69} nucleate above a Coulomb blockade threshold and form droplets resembling fluidic domain walls (soliton liquids), where interchain interactions or Josephson coupling between chains~\cite{70} prevent rampant thermal excitations;~\cite{52}  {2)} in the highest quality crystals the nucleation process is best described as coherent Josephson-like tunneling using a modified tunneling matrix element~\cite{1} that reflects the Zener probability; {3)} in these same materials, the time-evolution of complex order parameters, resembling probability amplitudes, can be described using the Schr\"{o}dinger equation as an emergent classical equation;~\cite{1,20} and {4)} both static (e.g., in ring experiments with magnetic flux) and dynamic (\textit{ac} response) vector and scalar potentials can couple to and/or modulate the phases of the complex order parameters.

\begin{figure}[h]	\centering		
\includegraphics[scale=.425]{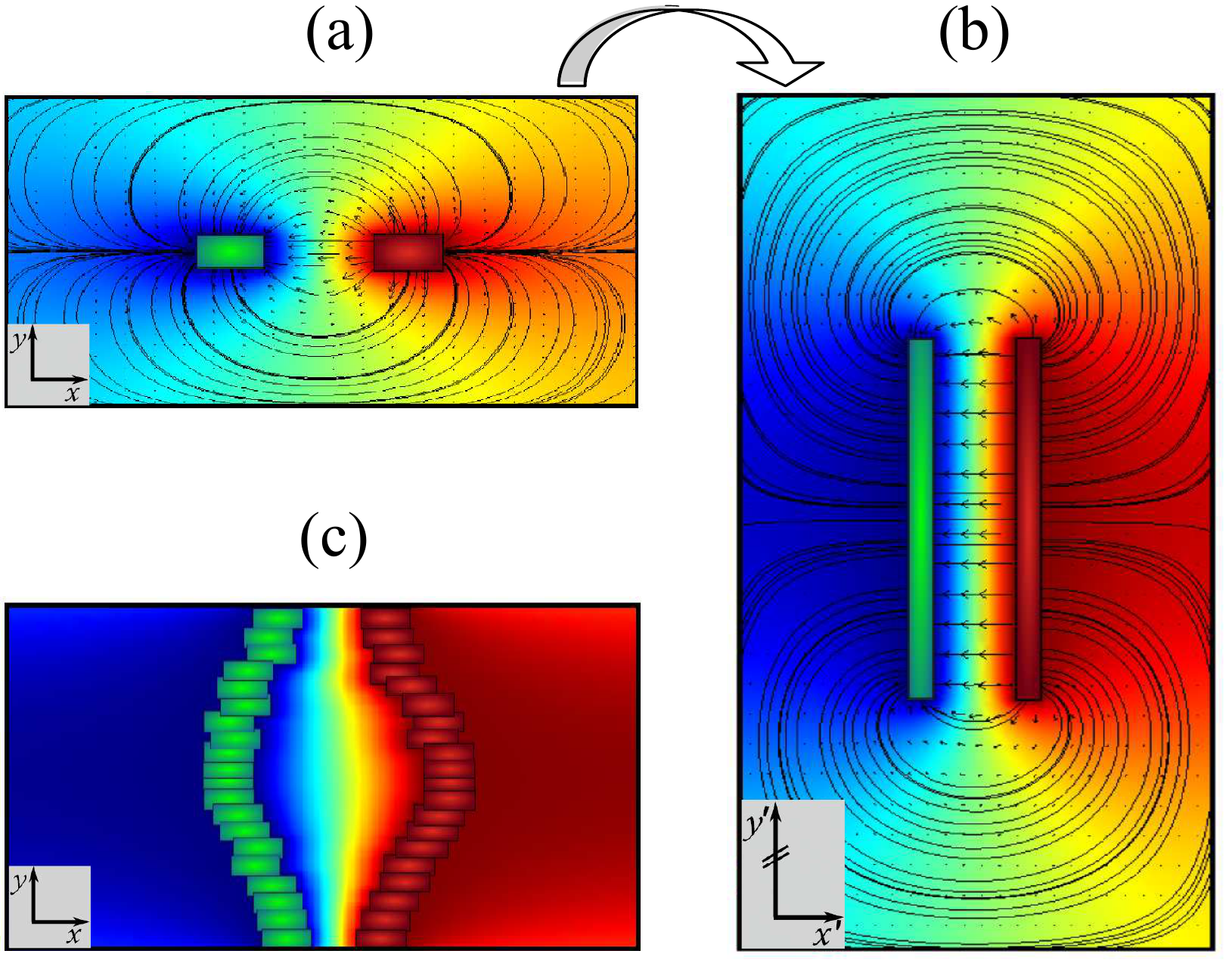}	
\caption{(a)~COMSOL simulation of electrostatic potential (red =~positive, blue =~negative) and field lines for an electric dipole consisting of dislocations represented as + and - rectangular charge distributions. (b)~COMSOL simulation for a similar pair with anisotropic dielectric constants, which resembles a parallel plate capacitor in rescaled coordinates. (c)~Aggregation of many dislocations into fluidic domain walls or droplets of soliton liquids, between which the bubble of lower energy or `true vacuum' grows as they are driven toward the contacts by the externally applied field.}	
\label{fig:Fig5}
\end{figure}

CDWs are often highly anisotropic, where the dielectric response, $\epsilon_{xx}$, along the chain direction is much greater than those, $\epsilon_{yy}$, and  $\epsilon_{zz}$ in the perpendicular directions. The degree of anisotropy affects the internal field $E^*$ generated by a dislocation pair (Fig.~\ref{fig:Fig5}(a), (b)) and, thus, the Coulomb blockade threshold field: $E_T=E^*/2$. One method of modeling this behavior (using COMSOL~\cite{71}) is to rescale the variables along the x-, y-, and z-directions by dividing by the relative dielectric constants: $x'=x/\epsilon_{xx}$, $y'=y/\epsilon_{yy}$, and $z'=z/\epsilon_{zz}$. This is seen starting with the Maxwell equation: $\nabla \cdot \textbf{D}=\rho$, where (using the summation convention): $\textbf{D}_i=\epsilon_0 \epsilon_{ij} E_j$. 
Here $\epsilon_{ij}$ is the relative dielectric tensor, which is diagonal with elements $\epsilon_{xx}$, $\epsilon_{yy}$, and $\epsilon_{zz}$, if the axes $i, j = x, y,$ and $z$ are along the principal crystallographic directions. Fig.~\ref{fig:Fig5}(b) illustrates the rescaled COMSOL simulations in 2-D, where the dislocation pair in rescaled coordinates looks like a parallel plate capacitor that produces an internal field $E^*=2e/2\epsilon A_{ch}= en_{ch}/\epsilon$, where $\epsilon=\epsilon_{xx}\epsilon_0$. This is within a factor of $1/2$ of the ideal value, $2en_{ch}/\epsilon$, for a fully condensed CDW. Figure~\ref{fig:Fig5}(c) shows the aggregation of many $2\pi$  dislocations of charge $2e$ into fluidic soliton droplets that move toward the contacts and allow the bubble of lower energy between them (or `true vacuum,' using the quantum field theory terminology) to grow. Other factors that can affect $E^*$ and $E_T$ include gate electrodes in CDW field- and current-effect transistors,~\cite{72,73,74,75,76,77} as well as screening by normal carriers.

\begin{figure}[h]	\centering		
\includegraphics[scale=.45]{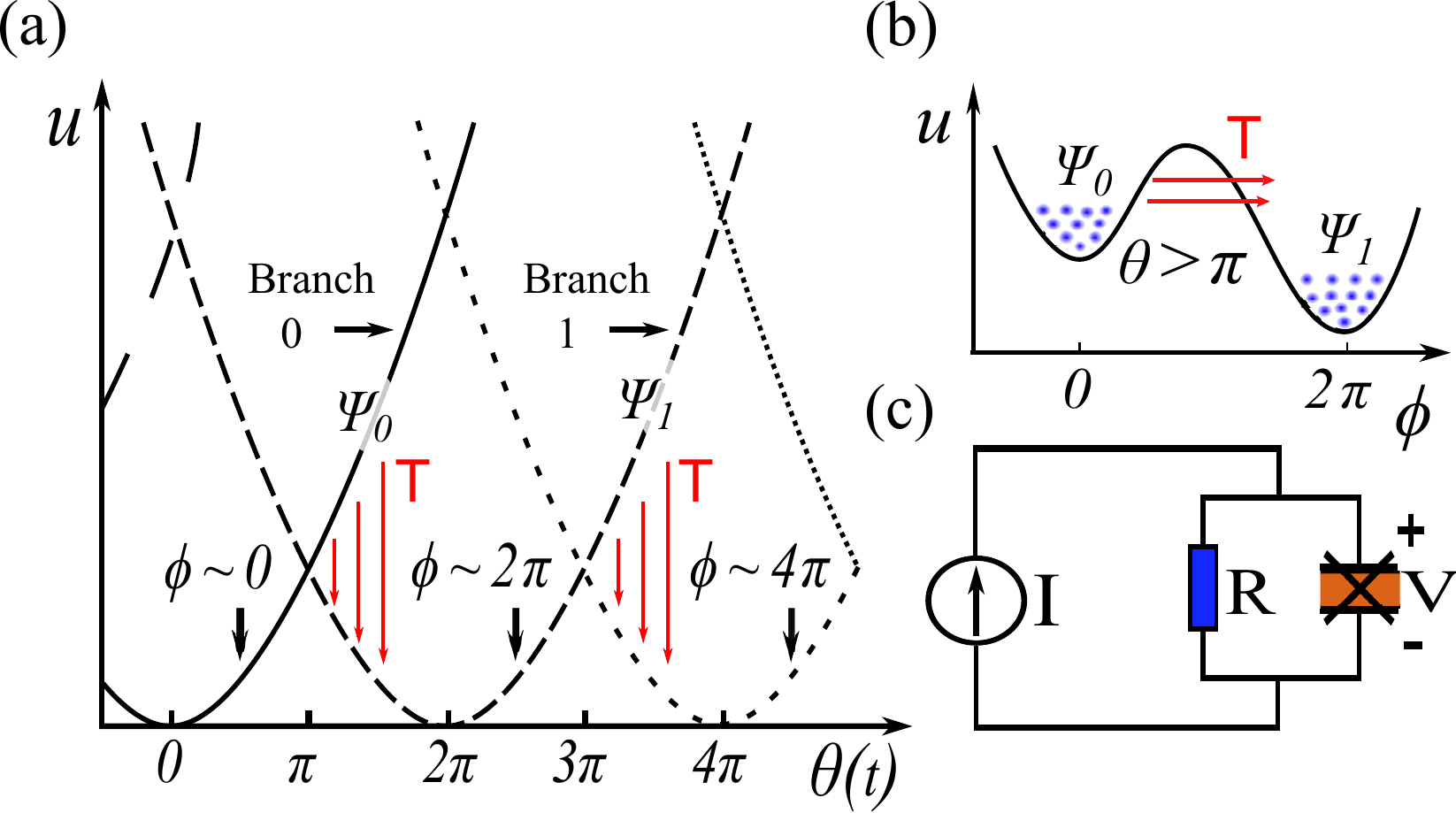}       
\caption{(a)~Potential energy \textit{vs.} $\theta$  for $\phi\sim 2\pi n$. (b)~$u$ \textit{vs.} $\phi$  when $\theta=2\pi E/E^* >\pi   (E>E_T)$ as the phases $\phi_k (x)$ tunnel coherently into the next well via the tunneling matrix element~$T$.  (c)~Time-correlated soliton tunneling model, consisting of a normal shunt resistance $R$ in parallel with the CDW, represented as a capacitive Coulomb blockade tunnel junction.}	
\label{fig:Fig6}
\end{figure}

The time-correlated soliton tunneling model,~\cite{1} which interprets CDW dynamics above threshold, borrows concepts from the theory of time-correlated single electron tunneling.~\cite{12,13} The electrostatic energy parabolas of Fig.~\ref{fig:Fig3} (also Fig.~\ref{fig:Fig6}(a)) are similar to the charging energies of a small-capacitance tunnel junction. According to this model, \textit{coherent voltage oscillations, narrow-band noise, and ac-dc interference effects come from these piecewise parabolic charging energy curves, and not from the shape of the periodic pinning potential.} The large normal carrier concentration in NbSe$_3$ due to incomplete Peierls gap formation leads to significant screening by normal carriers, which enhances the spatial uniformity of the CDW's dielectric response. This explains why highly coherent voltage oscillations, narrow-band noise peaks, and mode locking are often observed~\cite{63,78,79,80,81,82} in NbSe$_3$ crystals, even though the pinning comes from randomly distributed impurities.~\cite{29,30,55} Moreover, the piecewise parabolic curves also explain why the narrow-band noise spectra show such a rich array of harmonics.

The simplest model~\cite{1} treats the CDW as a capacitive Coulomb blockade tunnel junction in parallel with a shunt resistor $R$ (Fig.~\ref{fig:Fig6}(c)) due to normal uncondensed electrons. To model dynamics, the `vacuum angle'  $\theta$ is related to displacement charge $Q$ between contacts as: $\theta=2\pi Q/Q_0$, where $Q_0=2eN$ and $N$ is the number of parallel CDW chains. Advancing the phases of all chains by $2\pi  n$ creates multiple pairs of fluidic soliton domain walls that quickly reach the contacts. Similar to a capacitive tunnel junction the voltage is: $V=(Q-Q_0)/2C  = (Q_0/2\pi C)[\theta-2\pi n]$, where $C=\epsilon A/l$. More generally: $V=(Q_0/2\pi C)[\theta-\left\langle \phi\right\rangle]$, if $\left\langle \phi\right\rangle\neq 2\pi n$. The total current is: $I=I_n+I_{cdw}$, where $I_n=(Q_0/2\pi RC)[\theta-\left\langle \phi\right\rangle]$ is the normal current and $I_{cdw}=dQ/dt=(Q_0/2\pi)  d\theta/dt$ is the CDW current. (The latter includes capacitive displacement current but is identical to $(Q_0/2\pi)d\left\langle \phi\right\rangle/dt$ when time-averaged.) Defining $\omega\equiv 2 \pi I/Q_0$  and $\tau\equiv RC$ yields the following equation for the time evolution of  $\theta$: 

\begin{equation}\centering
	\frac{d\theta}{dt}=\omega-\frac{1}{\tau}\left[ \theta-\left\langle \phi\right\rangle\right].
	\label{eq:eq2}
\end{equation}

Since $\left\langle \phi\right\rangle$ advances in a jerky fashion, Eq.~(\ref{eq:eq2}) contains the elements needed to explain the observed voltage oscillations, narrow-band noise, etc. Within a unified framework it allows for at least three mechanisms by which $\left\langle \phi\right\rangle$ can evolve: (a)~coherent Josephson-like tunneling via a matrix element $T$, (b)~incoherent tunneling or thermal activation of solitons, modeled using probabilities instead of probability amplitudes, and (c)~classical depinning over the barrier, as in  Fig.~\ref{fig:Fig4}(b). Detailed studies of mechanisms (b) and (c) within this framework are potential topics of future investigation.  Equation ~(\ref{eq:eq2}) is important even in a classical picture, because it incorporates electrostatic effects and dissipative effects from the normal shunt resistance. Extensions beyond the single-domain model (e.g., using a coarse-grained network of CDW domains) would enable incorporation of random pinning and CDW deformability into this framework.

Feynman~\cite{20} (vol.~III, Ch.~21) provides a derivation of coherent Josephson tunneling, where the Schr\"{o}dinger equation is viewed as a `classical' equation to treat wavefunction-like order parameters coupled by a tunneling matrix element. We have developed~\cite{1} a similar method for the CDW to compute $\left\langle \phi(t)\right\rangle$ via the coherent tunneling mechanism (a).~It employs the Schr\"{o}dinger equation:

\begin{equation}\centering
	i\hbar\frac{\partial \psi_{0,1}}{\partial t}=U_{0,1}\psi_{0,1}+T\psi_{1,0}
	\label{eq:eq3}
\end{equation}
to compute the original and emerging probability amplitudes $\psi_0 (t)$  \& $\psi_1 (t)$ for the system to be on branches 0 and 1 in Fig.~\ref{fig:Fig6}(a) (more generally $\psi_j$  \& $\psi_{j+1}$) when coupled by the matrix element $T$. The model treats the amplitudes as complex order parameters:

\begin{equation}\centering
\psi_{0,1}=\sqrt{\rho_{0,1}}\exp \left[i\delta_{0,1}\right],
	\label{eq:eq4}
\end{equation}
where $\rho_{0,1}=N_{0,1}/N$ is the fraction of parallel chains on the respective branch. Advancing the CDW phases $\phi_k (x)$ of many chains by $2\pi$  (from one branch to the next in Fig.~\ref{fig:Fig6}) creates lower energy bubbles bounded by droplets of microscopic $2\pi$  solitons and antisolitons (somewhat delocalized as quantum solitons~\cite{14}) which form the new fluidic macrostate $\psi_1$. 

The microscopic quantum soliton energy per electron pair, $\Delta_\varphi$, can be estimated from the measured Zener field, $E_0\sim(\Delta_\varphi^ 2/\hbar \nu_0 e)$, typically $\sim$ 10~V/m. Using a phason velocity, $\nu_0 \sim  3\times 10^3$~m/s, yields $\Delta_\varphi \sim 5~\mu$eV, an extremely small value. However, the coupled macrostates have substantial condensation energies due to the many ($>10^9$) interacting parallel CDW chains.~\cite{46,52} The condensed solitons in the emerging macrostate are thus effectively trapped in soliton liquids, preventing thermal excitations except across the much larger Peierls gap. An analogy is provided by Josephson coupling between superimposed macrostates in 2-band superconductors,~\cite{83} where thermal excitations only occur across either BCS energy gap regardless of the energy difference between macrostates. One can also view bubbles of the CDW chains escaping out of the metastable well (Fig.~\ref{fig:Fig6}(b)) as being analogous to superfluid helium atoms quantum mechanically creeping out of a container. If the container rim is, for example, $d \sim$ 1~cm above the liquid surface, then the gravitational barrier per atom is $mgd \sim$ 4~neV, which is small compared to $kT$ even at 1~mK. Nevertheless, the helium atoms remain trapped in the superfluid, prevented by the condensation energy from thermally hopping out of the container even though they quantum mechanically creep over the rim in a collective fashion.

The driving force $F$ is the energy difference per unit length after one branch crosses another in Fig.~\ref{fig:Fig6}(a).~\cite{1} Using the analogy to pair production~\cite{49} and following Bardeen,~\cite{84,85} the tunneling matrix element $T$ is estimated to be:
\begin{equation}\centering
 T(F)=-4F\lambda \exp \left[-F_0/F\right],
	\label{eq:eq5}
\end{equation}
where $F_0 \sim \Delta_\varphi^2/\hbar \nu_0 $ and $\lambda$,  defined in ref.~\cite{1}, is comparable to the soliton width. As discussed above, $\nu_0=(m/M_F)^{1/2} \nu_F$ is the phason velocity, smaller than the Fermi velocity $\nu_F$ due to the large Fr\"{o}hlich mass ratio $M_F/m$.~\cite{46}  

\begin{figure}[h]	\centering		
\includegraphics[scale=.125]{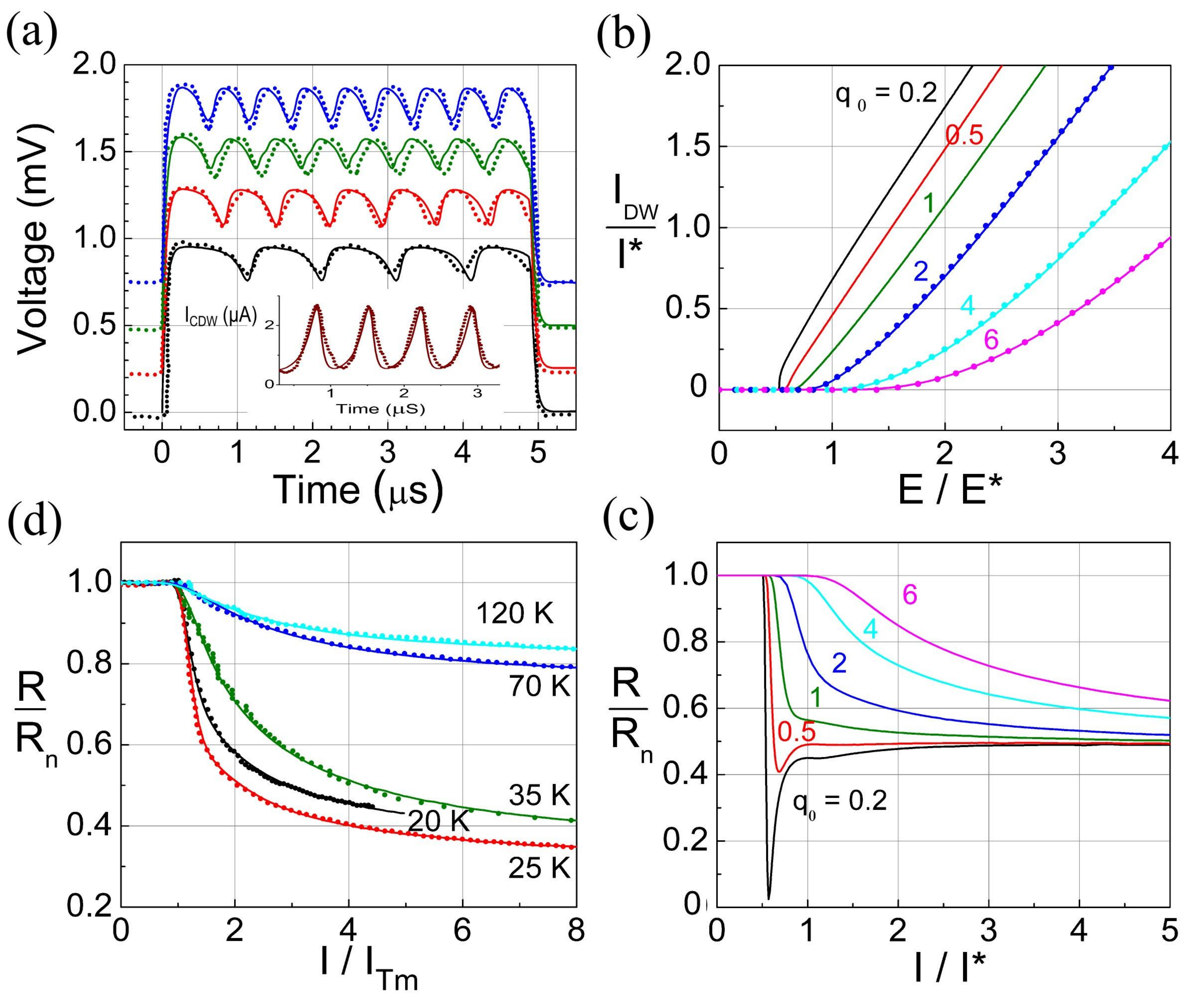}
\caption{(a)~Theoretical~\cite{1} (solid lines) \textit{vs.} experimental (dotted lines~\cite{78}) voltage oscillations (bottom to top, offset by 0, 0.25, 0.5, and 0.75 mV) of an NbSe$_3$ crystal at 52~K for current pulse amplitudes: 9.90~$\mu$A (black), 10.89~$\mu$A (red), 11.49~$\mu$A (green), \& 11.88~$\mu$A (blue).  Inset. CDW current, $I -  In$, \textit{vs.} time for the 10.89~$\mu$A pulse.  (b)~Simulated CDW current \textit{vs.} field for several $q_0 = F_0/2eE^*$. Dotted lines: Bardeen's modified Zener function.~\cite{16}  (c)~Simulated $R =$ \textit{dV/dI} \textit{vs.} current for several $q_0$, where $R_n$ is the normal resistance below threshold. (d)~Theoretical (solid lines) \textit{vs.} experimental (dotted lines) \textit{dV/dI} \textit{vs.} current for NbSe$_3$ (see~\cite{1} for parameters).}	
\label{fig:Fig7}
\end{figure}

Figure ~\ref{fig:Fig7}(a)~compares the simulations~\cite{1} with measured voltage oscillations~\cite{78} of NbSe$_3$ for rectangular current pulses. Except for the increasing pulse amplitudes, the same parameters are used for the entire family of theoretical plots (solid lines), which show unprecedented quantitative agreement with experiment. The model correctly captures the progression of non-sinusoidal shapes, ranging from rounded backward sawtooth behavior for the 9.90-$\mu$A current pulse to more symmetrical oscillations for higher pulse amplitudes. The inset to Fig.~\ref{fig:Fig7}(a)~shows the CDW current $(I_{cdw} = I - I_n)$ vs. time corresponding to the 10.89-$\mu$A pulse. This plot: 1) shows that a large fraction of the CDW current is oscillatory, and 2) captures the `flowing,' rather than abrupt tunneling, aspect of quantum transport. The \textit{I-V} and differential resistance curves are computed~\cite{1} by averaging over several cycles, with results shown in Figs.~\ref{fig:Fig7}(b)-(d). A range of behaviors are captured, ranging from rounded Zener-like behavior to more linear \textit{I-V} curves and \textit{dV/dI} curves with negative dips or wings, as seen in NbSe$_3$ crystals with fewer impurities.~\cite{63} The theoretical plots show outstanding quantitative agreement with experiment in Fig.~\ref{fig:Fig7}(d).

\section{Coupling of order parameter phases to vector and scalar potentials}\label{sec:4}
The order parameter phases $\delta_j$ of the branches in Eq.~(\ref{eq:eq4}) are not identical to the CDW phase $\phi$. The latter is the phase difference between CDW electron states separated by the nesting wavevector $Q = 2k_F$ (or electron and hole states if the CDW ground state is written in terms of a filled Fermi sea).  By contrast, $\delta \equiv \delta_{j+1}-\delta_j$ is likely related to the relative phases of nucleated soliton and antisoliton droplet order parameters. Since these carry charge, they couple directly to either a vector or scalar potential. The discussion here treats both the magnitudes and phases of the complex order parameters of Eq.~(\ref{eq:eq4}), $\psi_j=\sqrt{\rho_j}  \exp⁡[i\delta_j]$, as being classically robust for a system with enough parallel CDW chains. The simulations in section~\ref{sec:3}, of dc transport and rectangular current pulses, fix  $\delta$ at $ \pi /2$ in Eq.~(\ref{eq:eq3}),~\cite{1} which yields the maximum current in the Josephson current-phase relation. This section discusses coupling of static (magnetic field) and dynamic (\textit{ac} electric field) vector and scalar potentials to the phases $\delta_j$ in order to: a)~interpret the $h/2e$ quantum interference effects in CDW rings;~\cite{21,22} b)~better understand mixing and other \textit{ac} response experiments,~\cite{36,37,38} previously interpreted using photon-assisted tunneling (PAT) theory;~\cite{86} and c)~interpret large amplitude \textit{ac} experiments presented here. The ability to vary both magnitudes and phases of the macrostate amplitudes could eventually set the stage for development of future quantum computing devices, while better understanding of the ring experiments could enable new types of magnetic sensors.

In the CDW ring experiments~\cite{21,22} a static magnetic vector potential couples to the phases $\delta_j$ and leads to quantum interference between the amplitudes traversing the two branches of the ring (nucleated quantum solitons interfering with themselves). This can be visualized in terms of an extra phase shift  $\chi$ affecting the tunneling matrix elements of a two-domain model:

\begin{equation}
	T_{a,b}\rightarrow T\exp \left[\pm i\chi/2\right],
	\label{eq:eq6}
\end{equation}
one domain for each path, $a$ or $b$, along the ring. Here:

\begin{equation}
\chi=\frac{q}{\hbar}\oint \textbf{A}\cdot d\textbf{r}=2\pi\left[\Phi/\Phi_0\right],
\label{eq:eq7}
\end{equation}
where $\Phi_0=h/q$ and $q$ is either $e$ or $2e$. Summing the amplitudes then yields a modulation proportional to $|2T\cos [\pi\Phi/\Phi_0]|$. This simple two-domain picture gives the period $h/2e$ for the A-B oscillations provided we take $q=2e$. However, it is an oversimplification compared to the reported $\sim 10 \%$ modulation and rather disordered behavior in the A-B oscillations in the actual CDW rings,~\cite{21,22} suggesting the need to include many CDW domains with some degree of disorder. Moreover, \textit{ac} response experiments, discussed below, suggest a rather short tunneling distance. This further indicates the need to incorporate multiple domains,~\cite{87} which can be modeled as a network of many tunnel junctions in series.

The phases $\delta_i (t)$ of the macrostate order parameters in Eq.~(\ref{eq:eq4}) can be modulated by an oscillatory field that temporally evolves the scalar and/or vector potentials. The theory of photon-assisted tunneling (PAT)~\cite{86} enables predictions of tunnel junction response to combined \textit{dc} and \textit{ac} signals based on its \textit{dc} current-voltage (\textit{I-V}) characteristic. Oscillatory voltages modulate the relative energies and phases of wavefunctions on opposite sides of the tunnel junction. This generates various combinations of Bessel functions in the predicted responses, which reduce to finite differences of the \textit{I-V} curves in the small-signal limit.  A modification of PAT theory was previously adapted to interpret mixing and other CDW \textit{ac} response experiments~\cite{36,37,38} on TaS$_3$ and NbSe$_3$. These experiments show good agreement with PAT theory for small-amplitude signals.~\cite{36,37,38} The ``wavefunctions'' $\psi_j$ in the picture discussed here are viewed as classically robust complex order parameters so, in this regard, the term ``photon assisted tunneling'' (originally developed for single particle tunneling) may be a misnomer. However, some aspects of PAT theory may still apply, as suggested by previous experiments~\cite{36,38} and those discussed below.

Mixing experiments apply a signal of the form:
\begin{equation}
	V(t)=V_{dc}+ V_1\cos \omega_1t + V_2\cos \omega_2t
	\label{eq:eq8}
\end{equation}
and measure an induced response (e.g., with a lock-in amplifier): $\delta I(t)=\delta I_0 \cos[\omega_0 t+\varphi]$. The difference frequency is: $\omega_0=|\omega_2-\omega_1|$ for direct mixing and $\omega_0=|\omega_2-2\omega_1|$ for harmonic mixing. At low frequencies and amplitudes, the harmonic mixing response \textit{vs.} bias voltage $V_{dc}$ is proportional to the third derivative of the \textit{dc I-V} curve:

\begin{equation}
	\delta I_0(V_{dc})=\frac{1}{8}V_1^2V_2\left[\frac{d^3I_{dc}}{dV^3}\right]_{V=V_{dc}}.
	\label{eq:eq9}
\end{equation}
At finite frequencies, the third derivative gets replaced by a third finite difference~\cite{38} with a step size proportional to frequency but has a similar, albeit broadened, bias dependence. The harmonic mixing response at zero \textit{dc} bias voltage becomes significant for frequencies $\omega_0/2\pi$  of about 1~MHz and greater, and is found to be bias-independent below threshold.~\cite{38}  

Rather different behavior emerges when the \textit{dc} bias voltage $V_{dc}$ in Eq.~(\ref{eq:eq8}) is replaced by a large amplitude \textit{ac} `bias' voltage,
\begin{equation}
	V_{dc}\rightarrow V_{ac}\cos \omega t,
	\label{eq:eq10}
\end{equation}
and the harmonic mixing response $\delta I_0$ is plotted \textit{vs.} $V_{ac}$. When  $\omega$ is small, since harmonic mixing is an even function of \textit{dc} bias, $\delta I_0 (V_{ac})$ is just the time-averaged response \textit{vs.} bias voltage $\left\langle \delta I_0 (V_{bias}(t))\right\rangle$, which resembles a washed out third derivative. When $\omega /2\pi$ reaches about 50~kHz or higher, however, $\delta I_0 (V_{ac})$  resembles $\delta I_0 (V_{dc}-V_T )$ with an apparent threshold voltage collapsed to the origin.~\cite{38} The collapse of the \textit{I-V} curve is  likely caused by capacitive coupling due to the high CDW dielectric response, which suppresses the Coulomb blockade threshold at sufficiently high frequencies.

The most interesting behavior is expected to occur when  $\omega$ is at megahertz frequencies and higher. In the absence of coupling $T$, a macrostate of modulated effective energy $E_{0,1}$ corresponding to branch 0 or 1 in Fig.~\ref{fig:Fig6} would evolve as:
\begin{equation}
	\psi_{0,1}(t)=\psi_{0,1}(0)\exp \left[ -(i/\hbar) \int^{t}_{0} dt'E_{0,1}(t')\right].
	\label{eq:eq11}
\end{equation}
This modulation of energy levels by a time-varying voltage (scalar potential) is related to the scalar A-B effect,~\cite{26,28} where a time-varying scalar potential couples to the quantum-mechanical phase. (Future experiments could potentially study the combined effects of vector and \textit{dc} and/or oscillatory scalar potentials on CDW rings.)  Taking the charge to be $2e$, the voltage $V_{\ell}(t)=V_{\ell} \cos \omega t$ across a small domain of length $\ell$ modulates the energy $E_1(t)$ of state~1 relative to $E_0$, as: $\Delta E(t)=2eV_{\ell}\cos \omega t$. Macrostate $\psi_1$ then evolves relative to $\psi_0$ as:
\begin{eqnarray}
 \psi_1(t) &=& \psi_1(0)\exp \left[-iz \sin \omega t \right]      \nonumber \\
   &=& \psi_1(0) \sum^{\infty}_{n=-\infty} J_n(z) \exp \left[-in\omega t \right],
	\label{eq:eq12}
\end{eqnarray}
where $J_n (z)$ are Bessel functions and $z \equiv 2eV_\ell/\hbar \omega$. This effectively splits up the $\psi_1$ amplitude into many,

\begin{equation}
  \psi'_n=J_n(z)\psi_1,
	\label{eq:eq13}
\end{equation}
of virtual energy $E_n=n\hbar \omega$. 
Although these effective energies are extremely small \textit{per electron} (or \textit{per electron pair}), remember that the term inside the exponential on the RHS of Eq.~(\ref{eq:eq11}) is really a measure of the rate at which $\delta_{0,1}(t)$ evolves with time in a \textit{classically robust fashion}. By analogy, the ac Josephson effect is sometimes regarded as either due to the emission or absorption of photons of (extremely small) energy $\hbar\omega=2eV$ or to the classical time-evolution of the phase difference $\delta$, $\partial \delta/\partial t=2eV/\hbar$, across the junction.~\cite{19,20}
`Turning on' the tunneling matrix element $T$ enables it to couple states $\psi_0$  \& $\psi'_n$ of equal energy in Eq.~(\ref{eq:eq13}), any negative energy difference being balanced by the soliton pair energy.

Equation~(\ref{eq:eq13}) thus captures  essential features  of ``photon-assisted tunneling", where an initially occupied state can tunnel into an unoccupied virtual state of equal effective energy.  Recalling the relation between harmonic mixing and \textit{dc} bias voltage (Eq.~(\ref{eq:eq9}) and finite difference forms~\cite{38}), following PAT theory,~\cite{86,88} and noting that $J_{-n} (x)=(-1)^n J_n (x)$, the harmonic mixing response \textit{vs.} total voltage amplitude $V_{ac}$ between contacts would then be expected to be given by:


\begin{eqnarray}
 \delta I_0(V_{ac}) &=& J^{2}_{0} \left(\frac{V_{ac}}{\alpha\omega}\right) \delta I_0(V_{dc}=0) + \nonumber  \\
   && {}+2\sum^{\infty}_{n=1} J^{2}_{n} \left(\frac{V_{ac}}{\alpha\omega}\right) \delta I_0 (V_0=n\alpha\omega) 
	\label{eq:eq14}
\end{eqnarray}
where $V_0=V_{dc}-V_T$ due to the collapsed effective \textit{I-V} curve~\cite{38} at finite frequencies. The amplitudes and frequencies, $V_1$, $V_2$, $\omega_1$, and $\omega_2$, of the signals inducing the harmonic mixing response are fixed in these experiments.

The scaling parameter $\alpha$ in Eq.~(\ref{eq:eq14}) depends on the distance $L$ between contacts and an effective scaling length $\ell$, which relates the energy acquired by a particle of charge $e^*$ in an electric field $E$ to a quantum of energy $\hbar\omega$: $e^* V_{\ell}=e^* E\ell \leftrightarrow \hbar\omega$. Previously,~\cite{18} the effective charge was assumed to be a reduced by the Fr\"{o}hlich mass ratio, $e^*\sim \left[m/M_F\right]e \sim 10^{-3} e$, which yielded values of $\ell$ in the range 1.6-22~$\mu$m.~\cite{18}  Motivated by the recent CDW A-B ring experiments,~\cite{21,22} here we take the effective charge to be: $e^*=2e$, which reduces the estimated values of $\ell$ into the nanometer range. Further experiments are needed to determine the extent to which this effective charge is robust, since  some of the Fourier transformed A-B spectra in Fig.~\ref{fig:Fig2} of ref.~\cite{22} suggest multiple peaks at $h/e$, $h/2e$, and perhaps even $h/4e$, although the $h/2e$ peak appears dominant.  Charge $e$ could result, even for commensurability $M = 1$, from a decoupling of spin-up and spin-down CDW subbands in a $2\pi$ soliton dislocation or from a $\pi$ amplitude soliton,~\cite{65,66,67,68,69} while charge $4e$, for example, could result either from coupling of two parallel chains or, at sufficiently high bias voltages, from nucleation of $4\pi$ rather than $2\pi$ solitons.

In general, the nature of Zener-like tunneling through a tilted soliton gap may yield some degree of frequency- and/or field-dependence of $\ell$. Using $V_{ac}=(L/\ell) V_{\ell}$, one obtains the following scaling parameter (in this case taking $e^*=2e$): 

\begin{equation}
	\alpha=\frac{L}{\ell}\frac{\hbar}{2e}.
	\label{eq:eq15}
\end{equation}
Due to the properties of Bessel functions, the $J^{2}_{0} \left(V_{ac}/\alpha\omega\right)$ term in Eq.~(\ref{eq:eq14}) should initially dominate for small amplitudes, $V_{ac}$, while the remaining terms may become significant for larger $V_{ac}$.  Defining $\delta I_m \equiv \delta I_0 (V_{ac}=0)$ (usually a maximum), restricting $A_n (\omega)\equiv 2\delta I_0 (n\alpha\omega)/\delta I_m$ to be real, and keeping a finite number, $N$, of terms, the normalized theoretical harmonic mixing response can be approximated as:

\begin{equation}
	\frac{\delta I_o}{\delta I_m} \cong J^{2}_{0} \left( \frac{V_{ac}}{\alpha\omega}\right)+\sum^{N}_{n=1}A_n(\omega)J^{2}_{n}\left( \frac{V_{ac}}{\alpha\omega}\right).
	\label{eq:eq16}
\end{equation}

\begin{figure}[h]	\centering		
\includegraphics[scale=0.05]{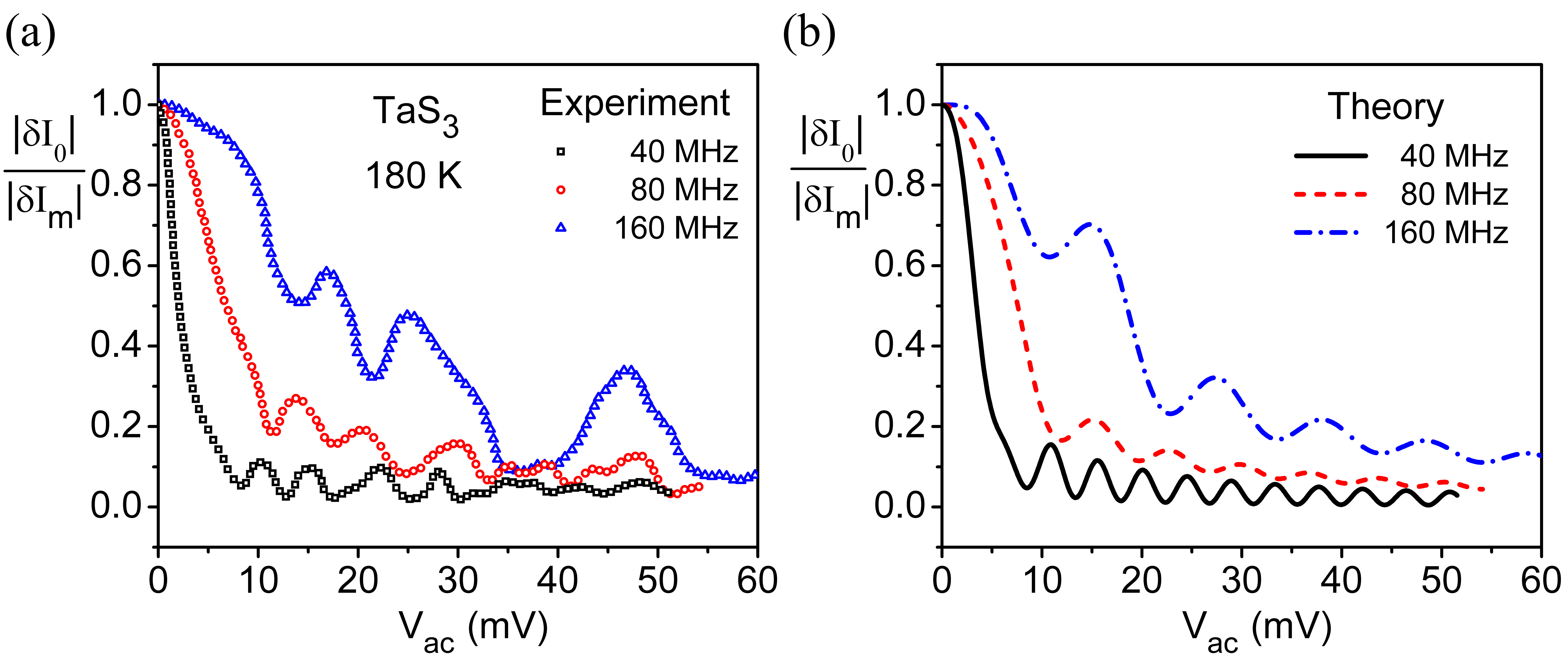}         
\caption{(a)~Normalized magnitude of harmonic mixing response ( $\omega_1/2\pi$  = 5~MHz,  $\omega_2/2\pi$  = 14~MHz,  $\omega_0/2\pi$  = 4~MHz) of a 0.1-mm long TaS$_3$ crystal, with a \textit{dc} threshold $V_T$ = 5~mV, \textit{vs.} \textit{ac} bias amplitude $V_{ac}$ at three different frequencies  $\omega/2\pi$  at 180~K. (b)~Theoretical plots using Eq.~(\ref{eq:eq16}) and the parameters shown in Table~\ref{tab:table1}.}	
\label{fig:Fig8}
\end{figure}

\begin{figure}[h]	\centering		
\includegraphics[scale=0.05]{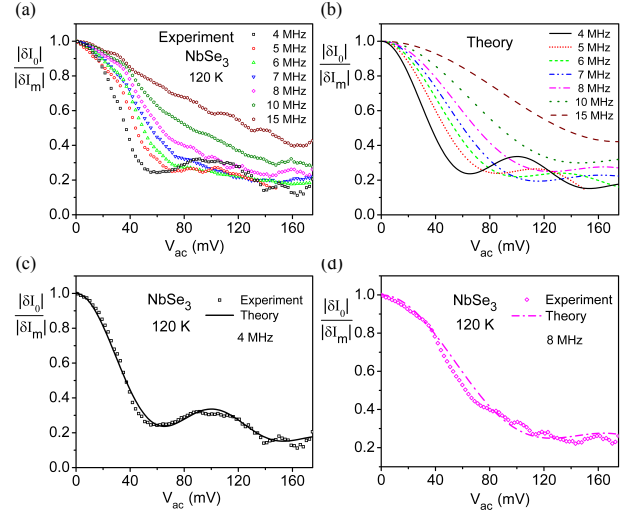}         	
\caption{(a)~Normalized magnitude of harmonic mixing response ( $\omega_1/2\pi$  = 1~MHz, $\omega_2/2\pi$  = 2.8~MHz,  $\omega_0/2\pi$  = 800~kHz) of an NbSe$_3$ crystal ($L$ = 5~mm), \textit{vs.} \textit{ac} bias amplitude $V_{ac}$ at several frequencies  $\omega/2\pi$  at 120~K. (b)~Theoretical plots using Eq.~(\ref{eq:eq16}) and the parameters shown in Table~\ref{tab:table2}.  (c) and (d)~Direct comparisons between theory and experiment for frequencies  $\omega/2\pi$ of 4~MHz and 8~MHz.}	
\label{fig:Fig9}
\end{figure}

Figures~\ref{fig:Fig8} and~\ref{fig:Fig9} show plots of normalized harmonic mixing responses $|\delta I_0|/|\delta I_m |$ \textit{vs.} \textit{ac} bias amplitude $V_{ac}$ for single crystals of TaS$_3$ and NbSe$_3$, as compared to Eq.~(\ref{eq:eq16}). Figure~\ref{fig:Fig8}(a) shows measured harmonic mixing responses of a TaS$_3$ crystal ($L$ = 0.1~mm) with a 5~mV threshold voltage at 180~K, for three different \textit{ac} bias frequencies $\omega$. Figure~\ref{fig:Fig8}(b) shows theoretical plots using Eq.~(\ref{eq:eq16}) and the parameters in Table~\ref{tab:table1}.  For this sample the effective scaling distance $\ell$, estimated using Eq.~(\ref{eq:eq15}) from the parameter $\alpha$, is found to be in the range 8-15~nm, or several CDW wavelengths.  The extremely small soliton energy gap per electron pair enables this distance to be longer than one would normally encounter in an ordinary tunnel junction.  We stress that the dc threshold effectively disappears at these frequencies,~\cite{36,38} making it unlikely that the behavior simply results from classically modulating the threshold field.

\begin{table}[h]
\caption{\label{tab:table1} Eq.~(\ref{eq:eq16}) parameters used for the Fig.~\ref{fig:Fig8}(b) theoretical plots.}
\begin{ruledtabular}
\begin{tabular}{cccccc}
\textrm{$\omega/2\pi$ (MHz)}&
\textrm{$\alpha$(V$\cdot$s)}&
\textrm{$A_1$}&
\textrm{$A_2$}&
\textrm{$A_3$}&
\textrm{$~A_4$}      \\
\colrule
~40&$5.5\times 10^{-12}$ &1.75 &0.25 &0.42 &-1.00 \\
~80&$4.2\times 10^{-12}$ &1.80 &1.40 &0.50 &-0.60 \\
160&$3.2\times 10^{-12}$ &2.00 &0.70 &1.57 &~1.44 \\
\end{tabular}
\end{ruledtabular}
\end{table}

Figure~\ref{fig:Fig9} shows measured harmonic mixing responses of an NbSe$_3$ crystal ($L$ = 5~mm) at 120~K, for several \textit{ac} bias frequencies  $\omega$. Figure~\ref{fig:Fig9}(a) shows experimental plots, while Fig.~\ref{fig:Fig9}(b) shows theoretical plots using Eq.~(\ref{eq:eq16}) and the parameters in Table~\ref{tab:table2}.  Figures~\ref{fig:Fig9}(c) and \ref{fig:Fig9}(d)  directly compare experiment with theory for \textit{ac} bias frequencies of 4~MHz and 8~MHz.  For this sample the effective scaling length $\ell$, estimated using Eq.~(\ref{eq:eq15}) from the scaling parameter  $\alpha$, is found to be 1.5~nm, or slightly greater than one CDW wavelength. Here, the shorter length $\ell$ may reflect a reduced effective mean free path length for the quantum solitons due the incomplete Peierls gap and large number of uncondensed normal carriers in NbSe$_3$. It will be interesting, in future studies, to determine whether the CDW wavelength represents an approximate lower bound on $\ell$.

\begin{table}[h]
\caption{\label{tab:table2} Eq.~(\ref{eq:eq16}) parameters used for the Fig.~\ref{fig:Fig9} theoretical plots.}
\begin{ruledtabular}
\begin{tabular}{ccccc}
\textrm{$\alpha=1.1\times 10^{-9}$(V$\cdot$s)}&
\textrm{$A_1$}&
\textrm{$A_2$}&
\textrm{$~A_3$}&
\textrm{$~A_4$}      \\
\colrule
~4~MHz&0.40 &0.65 &~0.15 &~0.55 \\
~5~MHz&0.45 &0.70 &-0.30 &~0.10 \\
~6~MHz&0.30 &1.00 &-1.40 &~1.90 \\
~7~MHz&0.20 &1.10 &-2.00 &~3.50\\
~8~MHz&0.20 &1.40 &-1.65 &~0.70 \\
10~MHz&0.15 &1.90 &-1.90 &-1.95 \\
15~MHz&0.01 &3.00 &~3.00 &1.00 \\
\end{tabular}
\end{ruledtabular}
\end{table}

The experiments reported here, as well as earlier mixing experiments at temperatures sometimes exceeding 200~K,~\cite{36, 37, 38} are consistent with the idea that oscillatory electric potentials modulate the phases of classically robust order parameters resembling macroscopic wavefunctions. Moreover, the experimental results are consistent with those of the CDW ring experiments,~\cite{21, 22} which demonstrate a significant degree of CDW quantum coherence. Collectively, the experiments support the hypothesis that either a vector or scalar potential couples to order parameter phases of CDW soliton condensates, and in some cases can lead to quantum interference. 

Further experimental and theoretical studies are warranted to enable the eventual development of a microscopic description of CDW transport. In particular, studies are needed to relate the variation of parameters $A_n$ in tables~\ref{tab:table1} and ~\ref{tab:table2} to the measured harmonic mixing response vs. frequency and bias voltage.  $A_1$ will usually be positive when mixing down to low or moderate frequencies since the harmonic mixing response is positive at low bias voltages. However, the remaining terms $A_n$ could either be positive or negative (higher frequencies sampling higher voltages via the voltage-frequency scaling) since the \textit{I-V} third derivative and harmonic mixing response become negative at certain bias voltages. A microscopic theory of CDW transport is ultimately needed to determine the extent to which previous~\cite{16,36,37,38} and current adaptations of PAT theory~\cite{86} are adequate or need modification, even for the quantum picture, and to which one can map the time evolution of the proposed complex order parameters onto a classical description.

\section{Discussion and Conclusion}
CDW transport is one of the few known cases of correlated transport of macroscopic numbers of electrons - the only known example of large-scale collective electron transport at human body temperatures.~\cite{4} This paper is highly transformative in that it challenges the classical sliding CDW paradigm that has dominated the field for over thirty years. Nevertheless, the quantum ideas discussed here can hardly be regarded as speculative. The evidence supporting quantum theory is so overwhelming, it can be considered a proven fact that electrons and all other known particles behave quantum mechanically. In 2000, the 100-year anniversary of Planck's black-body radiation paper,~\cite{89} Kleppner and Jackiw~\cite{90} pointed out that: ``Quantum theory is the most precisely tested and most successful theory in the history of science." Since then, aspects of quantum theory (the Pauli principle~\cite{91}) have been confirmed to within an accuracy of 6 x 10$^{-29}$. 

The classical behavior one observes on the macroscopic scale depends on the system and emerges from the behavior of large numbers of entangled quantum particles exhibiting wave-particle duality. The Schr\"{o}dinger equation can be regarded as the `classical' equation for superconducting condensates coupled through a thin insulator by Josephson tunneling (vol.~III, Ch.~21 of~\cite{20}). Similarly, the time-correlated soliton tunneling model discussed here treats the Schr\"{o}dinger equation as an emergent classical equation describing Josephson-coupled fluidic CDW macrostates. The simulations yield unprecedented quantitative agreement with coherent voltage oscillations and \textit{I-V} characteristics of NbSe$_3$ and also provide a natural interpretation for the quantum interference seen in the CDW ring experiments~\cite{21,22} and more complex interesting behavior seen in CDW harmonic mixing response.

Any further progress in understanding of CDW transport will require the scientific community to accept the fact that the CDW electron-phonon condensate behaves according to laws of quantum physics - the same quantum principles that govern every other system of particles in the universe. It is not necessarily true, \textit{a priori}, that quantum principles are consistent with the current dogma - that CDW electrons classically ``slide" according to Aristotle's linear velocity-force relation. Addressing the quantum behavior of CDWs, perhaps culminating in a microscopic theory of CDW transport and dynamics, would have enormous impact on this important branch of condensed matter physics. Additional areas of broad impact potentially include the boundary between CDWs and superconductivity, correlated electron-ion transport in biological systems, tunneling and `false vacuum decay' in quantum cosmology, a formally similar  $\theta=\pi$   instability for spontaneous \textit{CP} violation,~\cite{92} and a deeper understanding of quantum theory. 

Observation of quantum effects in NbS$_3$, which undergoes a Peierls transition well above room temperature,~\cite{4} would potentially lead to new devices such as magnetic sensors operating at room temperature. Understanding of the quantum behavior of solitons could lead to topologically robust (against decoherence) forms of quantum information processing, which would have major technological significance. 

Finally, the CDW may be one of the best systems yet to explore the boundary between the quantum world at the microscopic level and the emergent classical reality at the macroscopic scale. The `quantum-classical' paradigm proposed here and in our previous paper~\cite{1} provides further impetus for exploring this boundary, as do the recent CDW ring~\cite{21, 22} and related experiments that still await a complete microscopic description.

\begin{acknowledgments}
This work was supported by the State of Texas through the Texas Center for Superconductivity at the University of Houston.
\end{acknowledgments}

\end{document}